# 30-meter Land Surface Temperature from Landsat via Progressive Self-Training Downscaling

***Huanfeng Shen*[a,b,c], *Chan Li*[a], *Menghui Jiang*[a*], *Penghai Wu*[d], *Guanhao Zhang*[a], *Tian Xie*[a]**

[a] *School of Resource and Environmental Sciences, Wuhan University, Wuhan 430079, China*

[b] *Key Laboratory of Geographic Information System of Ministry of Education, Wuhan 430079, China*

[c] *Key Laboratory of Digital Cartography and Land Information Application of the Ministry of Natural Resources, Wuhan 430079, China*

[d] *Anhui Province Key Laboratory of Wetland Ecosystem Protection and Restoration, Anhui University, Hefei, Anhui 230601, China*

***Corresponding author.**

*E-mail address:* jiangmenghui@whu.edu.cn (M.Jiang)

**Abstract：** Land surface temperature (LST) is a critical parameter for characterizing surface energy balance and hydrothermal processes. While Landsat provides invaluable LST observations at medium spatial resolution for over 40 years, its native spatial resolution of thermal bands (e.g., 100 m) remains insufficient compared to its 30 m optical bands, failing to meet the demands of fine-scale studies. To address this issues, this study proposes a progressive self-training framework for downscaling Landsat LST to 30 m without relying on fine-scale ground truth, while maintaining minimal data dependence. The framework progressively optimizes a cross-modal fusion network to refine thermal details in a coarse-to-fine manner, characterized by one pre-training and two fine-tuning stages. This network leverages a cross-modal dictionary attention mechanism combined with a hybrid CNN-Transformer architecture to effectively extract complementary information between LST and auxiliary variables across multiple spatial resolutions. In parallel, a frequency-domain guidance strategy, implemented at both the network structure and loss-function levels, is introduced to enhance structural fidelity and suppress artifact accumulation during progressive refinement. Spatial validation against SDGSAT-1 30 m LST and temporal validation using in situ measurements confirm its reliability and accuracy, with both station-averaged MAE and RMSE outperforming the official cubic product by approximately 0.4 K. Further performance comparison experiments demonstrate that the proposed framework consistently reconstructs coherent fine-scale thermal patterns while preserving spatial heterogeneity. Multi spatial resolution evaluations and ablation studies verify the effectiveness of the proposed strategy and network design. Overall, the framework provides a stable pathway for enhancing the spatial resolution of Landsat LST, providing fine-resolution data support for fine-scale surface process studies and localized environmental monitoring.

## 1. Introduction

Land surface temperature (LST) governs energy, water, and carbon exchanges at the land-atmosphere interface, acting as a crucial indicator for quantifying surface-atmosphere interactions and their dynamics across regional to global scales (Myneni et al., 1997, Sellers et al., 1997). Satellite thermal infrared (TIR) remote sensing serves as the most effective means for acquiring LST with extensive spatial coverage(Li et al., 2013). Among various satellite missions, Landsat series occupies an unparalleled position in TIR monitoring due to its medium-resolution and unprecedented long-term consistent record for over 40 years (Wulder et al., 2019). However, the native spatial resolution of Landsat TIR data (e.g., 100 m for Landsat 8/9) still struggles to meet the demands of precision-demanding applications(Mukherjee et al., 2014, Hutengs and Vohland. 2016), such as urban thermal environment assessment (Weng et al., 2004), precision agriculture monitoring (Anderson et al., 2012), and fine-scale surface energy balance modeling (Kustas et al., 2009). Therefore, the issue of enhancing the spatial resolution of Landsat LST has attracted widespread attention (Zhan et al., 2013).

Downscaling is an essential approach for obtaining fine-resolution surface parameters by incorporating fine-resolution auxiliary information to guide spatial disaggregation (Jiang et al., 2024, Jing et al., 2024). According to the source of fine-resolution auxiliary information, prevailing LST downscaling methodologies can be broadly categorized into spatiotemporal fusion methods and multi-parameter fusion-based approaches (Jiang et al., 2024, Jing et al., 2024). Spatiotemporal fusion methods leverage complementary characteristics of satellite with fine spatial but coarse temporal resolution (e.g., Landsat) and those with fine temporal but coarse spatial resolution (e.g., MODIS) (Yin et al., 2020, Wu et al., 2022), and learn linear or nonlinear mappings between them to generate products with concurrently fine spatial and high temporal resolution (Hu et al., 2022, Li et al., 2023). The technical LST spatiotemporal fusion approaches has evolved from weight-based filtering methods that combine neighborhood information (Zhu et al., 2010), to unmixing and sparse-coding approaches leveraging dictionary learning (Song et al., 2012), and more recently, to deep learning frameworks that capture complex spatiotemporal dependencies and structural details (Tan et al., 2018). However, these methods rely heavily on the availability of high-quality, clear-sky reference image pairs acquired at proximate dates. Such strict requirements are often difficult to satisfy in cloud-prone regions or seasons, thereby limiting their practical applicability for continuous environmental monitoring.

Multi-parameter fusion-based methods enhance spatial resolution by establishing relationship between coarse-resolution LST and fine-resolution surface variables (such as vegetation indices) (Agam et al., 2007, Tang et al., 2021). These methods are primarily subdivided into empirical statistical approaches, physically-based algorithms and machine learning methods. Among them, empirical statistical approaches capture statistical relationships between LST and surface variables (Wu and Li. 2019, Dong et al., 2020), but they are limited in generalization due to reliance on predefined functions that struggle to represent heterogeneous and nonlinear interactions (Zawadzka et al., 2020, Liang et al., 2023). Physically-based algorithms derive LST from thermal radiative transfer or energy balance principles for high interpretability (Dominguez et al., 2011, Li et al., 2021) but suffer from poor spatial generalizability due to their sensitivity to regionally variable parameters (Bastiaanssen et al., 1998, Su and sciences. 2002). Machine learning methods leverage abundant samples to automatically learn complex nonlinear mappings between LST and auxiliary variables. These include early simple machine learning models like random forest (Li et al., 2021) and support vector machine(Xu et al., 2021), which outperform empirical regressions in capturing nonlinearity, as well as recent deep learning that gained prominence due to capabilities for automatic hierarchical feature extraction from high-dimensional data (Zhang et al., 2021, Hu et al., 2023, Li et al., 2025).

Despite these advances, the current LST downscaling research exhibits imbalance. While methodologies for kilometer-scale sensors (e.g., MODIS) are well-established (Hutengs and Vohland. 2016, Firozjaei et al., 2024), approaches tailored for medium resolution LST from sensor like Landsat remain scarce. Furthermore, even among the sparse medium resolution LST downscaling research, the complementary cross-modal features between LST and heterogeneous auxiliary variables are often insufficiently exploited through simple algorithmic designs, leading to unstable model performance, particularly in areas with high surface heterogeneity. Beyond these limitations, the field is most fundamentally constrained by the absence of fine-resolution LST ground truth. Consequently, the few existing Landsat LST downscaling studies resort to multi-parameter fusion-based approaches built upon the scale invariance assumption, presuming that the relationship between LST and auxiliary variables remains consistent across spatial resolutions (Dong et al., 2020). This relationship is established at the coarse Landsat resolution and applied to finer auxiliary variables to estimate finer-resolution LST. While this framework has achieved some success (Zhan et al., 2013, Guo et al., 2024) , it faces intrinsic limitations stemming from the frequent violation of the core scale invariance assumption. Robust evidence demonstrates that the statistical relationships between LST and auxiliary variables are scale-dependent, varying with spatial resolution (Bindhu et al., 2013, Duan et al., 2016) or spatial extent (Hu et al., 2023). Neglecting the scale effects is a primary source of accuracy limitation and uncertainty.

Therefore, this study proposes a dedicated progressive self-training learning framework to mitigate scale effects, which achieves the gradual downscaling of fine-scale thermal features and produces 30 m Landsat LST without relying on real fine-resolution ground truth. The main contributions are summarized as follows. We propose a progressive self-training framework that reformulates Landsat LST downscaling into a multi-stage, coarse-to-fine learning process. LST pseudo-labels generated from the previous coarse-resolution model are used to guide the training of the subsequent fine-resolution model, allowing the model to gradually adapt to fine-resolution feature patterns without relying on real fine-resolution labels. The proposed framework is designed to ensure minimal data dependence and strong operational feasibility. It relies solely on Landsat observations and readily accessible Digital Elevation Model (DEM) data, without introducing external high-resolution auxiliary datasets. To effectively integrate cross-modal features, we develop a representationally powerful cross-modal fusion network combining a cross-modal dictionary attention fusion module with a hybrid CNN-Transformer architecture to extract and integrate local and global complementary information between LST and auxiliary variables. Furthermore, a dual frequency-domain guidance mechanism, integrating internal high-frequency preservation blocks and external structural loss constraints, is employed in the network to model fine textures while suppressing artifacts inherent in LST pseudo-labels.

The remainder of this paper is organized as follows. Section 2 describes the study area and data. Section 3 introduces the proposed method. Section 4 presents the results and evaluation. Sections 5 and 6 provide discussion and conclusions, respectively.

## 2. Study area and data

### *2.1 Study area*

To construct a robust and representative dataset for Landsat LST downscaling, this study conducted sampling across China. Fig. 1 illustrates the geographical locations of the study and validation area. Six representative regions (Fig. 1 (a–f)) were selected to capture a wide spectrum of surface conditions and thermal environments. These regions encompass diverse land cover types, including forested mountains, agricultural plains, urban impervious surfaces, barren land and snow/ice. This high spatial heterogeneity, driven by diverse topography and land cover, provides an ideal study ground for evaluating the generalization ability and stability of LST downscaling models across different surface types. To further validate the downscaling results, this study selects the Heihe River Basin as the validation area (Fig. 1 (g)), from which seven ground stations are chosen for evaluation.

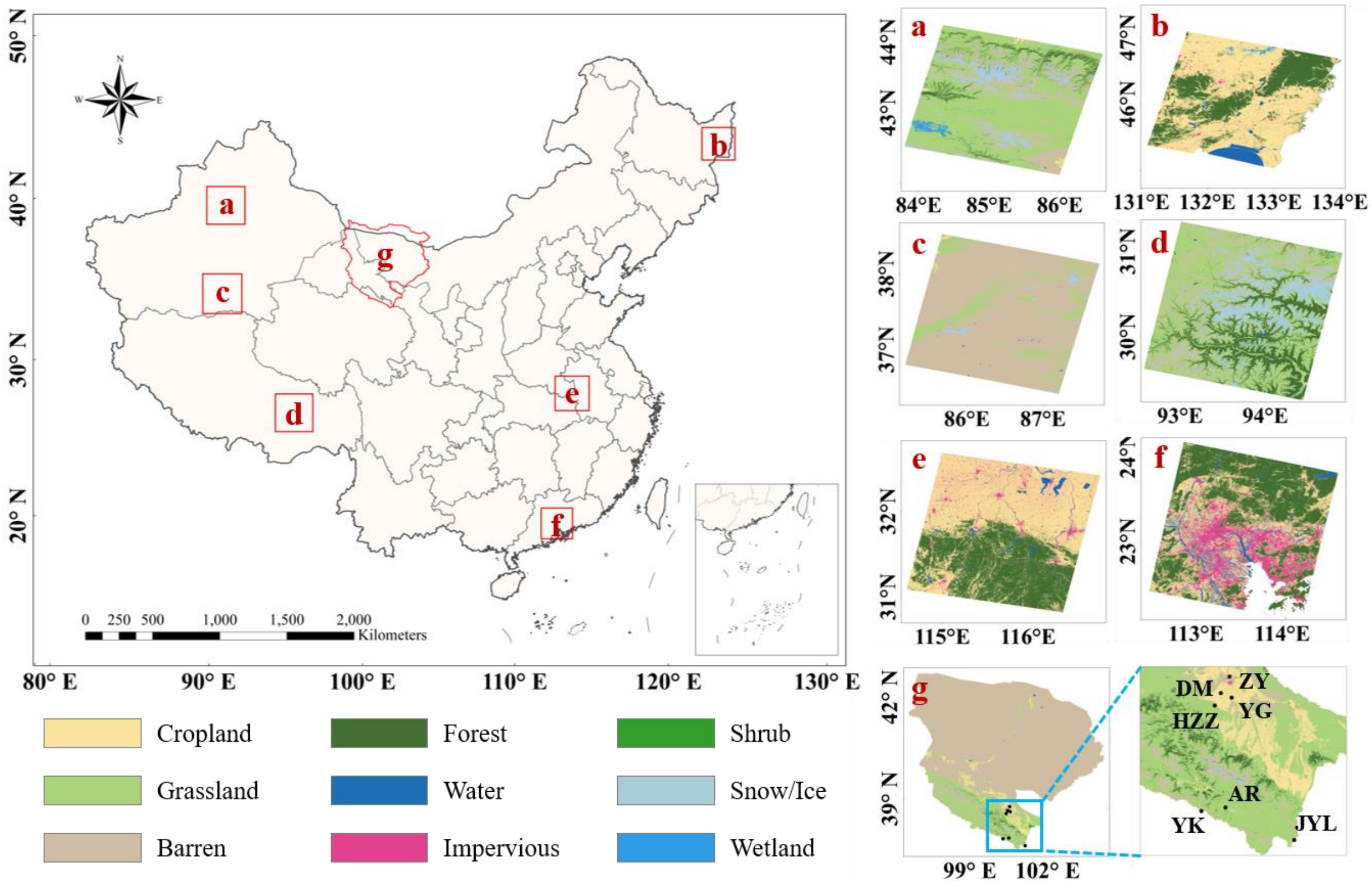

Fig. 1 Study area and station locations, where 7 station locations are marked with solid black dots.

*2.2 Study Data*

In this study, two main types of data were utilized: LST and auxiliary surface variables. Landsat 8/9 LST products serve as the target for downscaling, while in situ station observations and fine-resolution SDGSAT-1 TIR data are used for validation. Detailed descriptions of the datasets are provided in Table 1.

**Table 1**

Details of experimental data

<table>
<tr><th>Data</th><th>Source</th><th>Native Resolution (m)</th><th>Resampled Resolution (m)</th><th>Date</th><th>Usage</th></tr>
<tr><td rowspan="3">LST</td><td>Landsat 8/9 TIRS/TIRS-2</td><td>100 m</td><td>240 m, 120 m</td><td>2021.12.29, 2022.03.03, 2022.03.11, 2022.04.04, 2022.07.05, 2022.09.03, 2023.04.01, 2023.11.19</td><td>Input</td></tr>
<tr><td>SDGSAT-1 TIS</td><td>30m</td><td>——</td><td>2022.03.11</td><td>Validation</td></tr>
<tr><td>In situ</td><td>Point</td><td>——</td><td>2022.01.01-2023.12.31</td><td>Validation</td></tr>
<tr><td rowspan="2">Auxiliary variables</td><td>Landsat 8/9 OLI/OLI-2 NDVI, NDWI, NDBI</td><td rowspan="2">30m</td><td rowspan="2">120 m, 60 m, 30 m</td><td rowspan="2">2021.12.29, 2022.03.03, 2022.03.11, 2022.04.04, 2022.07.05, 2022.09.03, 2023.04.01, 2023.11.19</td><td rowspan="2">Input</td></tr>
<tr><td>SRTM1 DEM</td></tr>
</table>

(1) Landsat LST

Landsat 8/9 Collection 2 Level-2 (C2 L2) products (Cook et al., 2014, Cook. 2014, Landsat. 2021), released by the United States Geological Survey (USGS), were used as the primary LST source. The C2 L2 products are atmospherically corrected and radiometrically calibrated, providing high-quality surface measurements. LST was derived from the Landsat 8/9 Band 10 (TIRS/TIRS-2), which was selected over Band 11 due to its higher signal-to-noise ratio and substantially reduced stray-light interference. Although the USGS resamples TIR data to 30 m via cubic convolution to match the optical bands, the native acquisition resolution remains 100 m.

(2) Auxiliary surface variables

The auxiliary surface variables were primarily derived from Landsat 8/9 OLI/OLI-2 surface reflectance, including the Normalized Difference Vegetation Index (NDVI), Normalized Difference Water Index (NDWI), and Normalized Difference Built-up Index (NDBI), which are all closely associated with surface thermal environment (Chen et al., 2024). Additionally, the DEM from the Shuttle Radar Topography Mission (SRTM) was integrated to provide topographic information. These four variables served as fine-resolution inputs alongside the coarse-resolution LST to construct multi-resolution training samples for the downscaling model.

(3) In situ data

To quantitatively evaluate the accuracy of the downscaled LST, in situ measurements from the Heihe River Basin were obtained via the National Tibetan Plateau Data Center (https://data.tpdc.ac.cn/)(Wang et al., 2005, Liu et al., 2018, Che et al., 2019, Liu et al., 2023). Detailed station information is summarized in Table 2. (Cook. 2014). Two superstations and five conventional weather stations were selected: A'rou Superstation (AR), Daman Superstation (DM), Jingyangling station (JYL), Heihe Remote Sensing Station (HH), Huazhaizi Desert Steppe Station (HZZ), Yakou Station (YK), and Zhangye Wetland Station (ZY). These stations are equipped with four-component radiometers that record surface upwelling and atmospheric downwelling longwave radiation at 10-minute intervals. The in-situ LST $T_s$ is retrieved using the Stefan–Boltzmann law, as given by the following equation (Duan et al., 2019):

$$T_s = \left[\frac{L^{\uparrow} - (1-\varepsilon)L^{\downarrow}}{\varepsilon \cdot \sigma}\right]^{1/4} \tag{1}$$

where $L^{\uparrow}$ and $L^{\downarrow}$ represent the surface upwelling and atmospheric downwelling longwave radiation, respectively, and $\sigma$ is the Stefan–Boltzmann constant ($5.67 \times 10^{-8}\, W \cdot m^{-2} \cdot K^{-4}$). The surface broadband emissivity was estimated using the empirical relationship below

(Wang et al., 2005):

$$\varepsilon = 0.2122 \cdot \varepsilon_{29} + 0.3859 \cdot \varepsilon_{31} + 0.4029 \cdot \varepsilon_{32} \quad (2)$$

where $\varepsilon_{29}$, $\varepsilon_{31}$ and $\varepsilon_{32}$ are the narrowband emissivities corresponding to MODIS bands 29, 31, and 32, respectively.

**Table 2**

Detailed information of in situs sites

| Abbreviation | Latitude (°N) | Longitude (°E) | Elevation (m) | Instrument height (m) | Landscape |
|---|---|---|---|---|---|
| AR | 38.04 | 100.46 | 3033 | 5 | Subalpine mountain meadow |
| DM | 38.85 | 100.37 | 1556 | 12 | Maize |
| HH | 38.82 | 100.47 | 1560 | 1.5 | Cultivated grassland |
| HZZ | 38.76 | 100.32 | 1731 | 6 | Kalifan Kalidium piedmont desert |
| JYL | 37.83 | 101.11 | 3750 | 6 | Alpine meadow |
| YK | 38.01 | 100.24 | 4148 | 6 | Alpine meadow |
| ZY | 38.97 | 100.44 | 1460 | 6 | Reed wetland |

(4) SDGSAT-1 TIR data

For further independent spatial validation, 30 m LST derived from the SDGSAT-1 satellite Thermal Infrared Spectrometer (TIS) were employed. SDGSAT-1 provides three TIR bands with 30 m spatial resolution and strong thermal radiation detection capability. In this study, LST was retrieved from the TIS data using the split-window-driven temperature-and-emissivity separation (SWDTES) method (Wang et al., 2024). The 30 m SDGSAT-1 LST serves as a reference for evaluating the spatial structural of the downscaled Landsat LST.

(5) Data processing

To ensure spatial consistency across all datasets, a standardized preprocessing procedure was applied. All datasets were first reprojected to the WGS84 coordinate system to ensure geographic consistency. Considering the resolution requirements of the proposed progressive learning strategy, Landsat 8/9 LST were resampled to 240 m and 120 m spatial resolutions (rather than the native 100 m spatial resolution). The four auxiliary datasets were resampled to 120 m, 60 m, and 30 m spatial resolutions, to generate multi-resolution training pairs for coarse-to-fine supervision.

## 3. Methods

To address the scarcity of fine-resolution LST products and the reliance of existing models on scale invariance assumption, this study proposes a progressive self-training framework for Landsat LST downscaling through cross-modal fusion. The framework adopts a multi-stage strategy of initial pre-training followed by pseudo-label-guided fine-tuning to progressively learn scale-consistent LST mappings from 120 m to 60 m and further to 30 m. The cross-modal fusion network combines dictionary attention mechanism and CNN-Transformer modules to effectively extract complementary information between coarse-resolution LST and fine-resolution auxiliary variables, while embedded high-frequency preservation block and dedicated high-frequency structural loss to preserve spatial details at each resolution stage. These designs jointly enable accurate and stable Landsat LST downscaling despite the absence of real fine-resolution labels. The following sections describe the progressive self-training framework, cross-modal fusion network, and loss formulation in detail.

### *3.1 Progressive Self-Training Framework*

The overall framework, as illustrated in Fig. 2, constructs a cross-scale learning chain through three progressive stages: pre-training (240 m → 120 m), first fine-tuning (120 m → 60 m) and second fine-tuning (60 m → 30 m). It starts with training a cross-modal fusion model at a coarse resolution to establish a stable mapping between LST and auxiliary variables. The previous model is then applied to generate LST pseudo-labels for the next finer resolution. In each fine-tuning stage, the downsampled pseudo-labels, together with fine-resolution auxiliary variables, are used as inputs, while the original pseudo-labels provide weak supervisory guidance. This stage-wise self-training approach enables the model to gradually learn fine-scale thermal patterns, ultimately achieving accurate Landsat LST downscaling to 30 m.

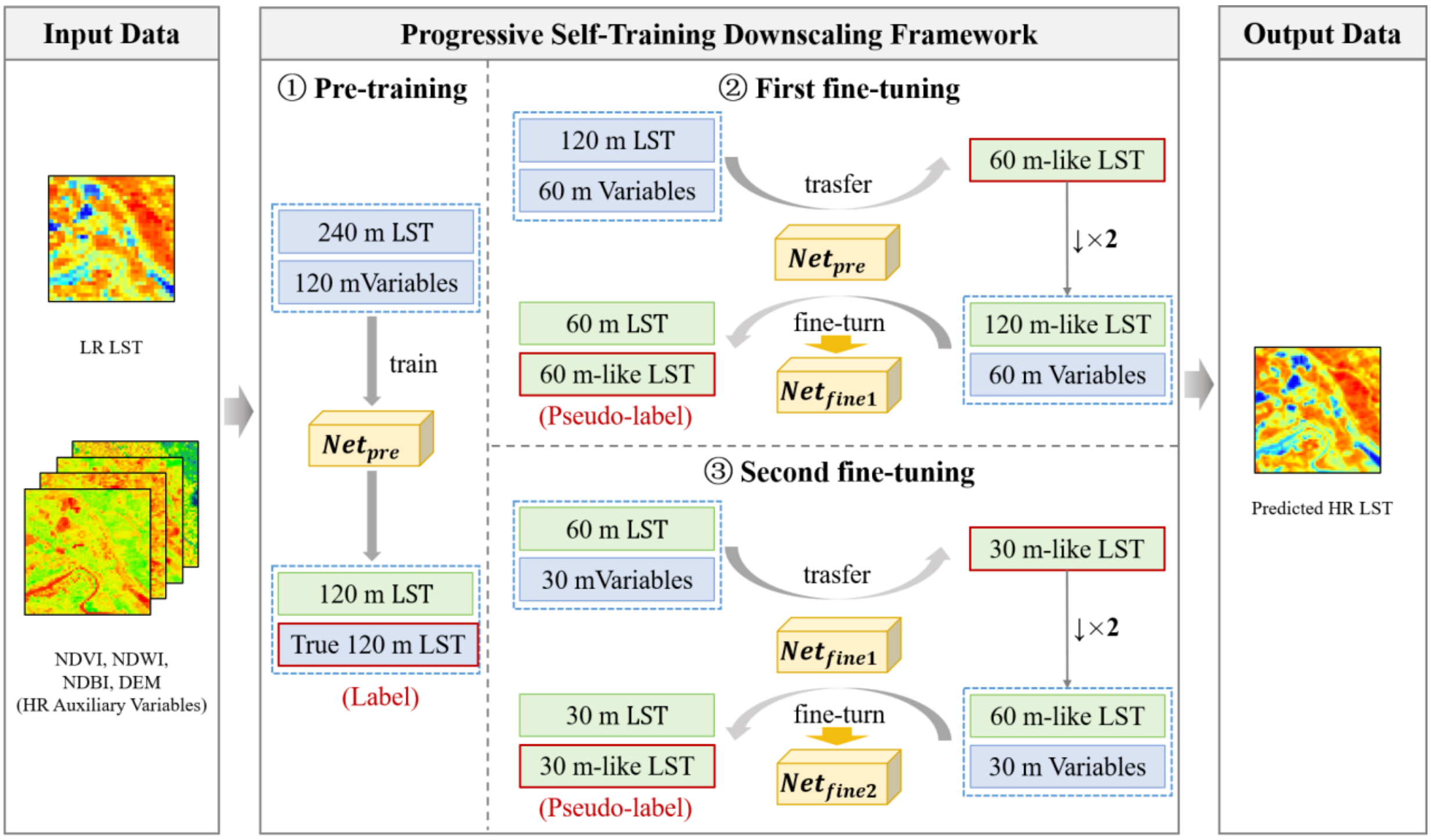

Fig. 2 Flowchart of the progressive self-training downscaling framework, where LR and HR denote the low- and high-resolution versions in the downscaling process, respectively.

(1) Pre-training stage (240 m → 120 m):

This foundational stage aims to learn a robust initial mapping between coarse-resolution LST and fine-resolution auxiliary variables, which can be formally expressed as:

$$T_{pred}^{120} = Net_{pre}(T^{240}, Aux^{120}) \tag{3}$$

where, $Net_{pre}$ denotes the pre-trained model; $T^{240}$ represents the input 240 m LST; $Aux^{120}$ refers to the 120 m auxiliary surface variables, include $NDVI^{120}, NDWI^{120}, NDBI^{120}, DEM^{120}$; and $T_{pred}^{120}$ indicates the downscaling result of the pre-trained model. The available satellite-retrieved 120 m LST can serve as label for network training in this stage.

(2) First fine-tuning stage (120 m → 60 m):

After the pre-training, the framework proceeds to the first fine-tuning stage. The pre-trained model $Net_{pre}$ is applied to 120 m LST with 60 m auxiliary variables to generate the

60 m LST pseudo-label:

$$T_{pl}^{60} = Net_{pre}(T^{120}, Aux^{60}) \tag{4}$$

where $T^{120}$ is the satellite-retrieved 120 m LST; $Aux^{60}$ denotes the 60 m auxiliary variables; and $T_{pl}^{60}$ is the 60 m pseudo-label. Using the downsampled version of the pseudo-label ($T_{de}^{120}$) and 60 m auxiliary variables, the model is further fine-tuned to generate the 60 m LST, expressed as:

$$T_{pred}^{60} = Net_{fine1}(T_{de}^{120}, Aux^{60}) \tag{5}$$

where $Net_{fine1}$ is the first fine-tuned model initialized with the weights of $Net_{pre}$; and $T_{pred}^{60}$ denotes the predicted 60 m LST. At this stage, the pseudo-label and the fine-resolution auxiliary variable serve as supervisory signals.

(3) Second fine-tuning stage (60 m → 30 m):

Following the same progressive strategy, the second fine-tuning stage refines the first fine-tuned model at the target 30 m resolution. The model $Net_{fine1}$ is applied to the 60 m prediction together with 30 m auxiliary variables to generate the 30 m LST pseudo-label:

$$T_{pl}^{30} = Net_{fine1}\left(T_{pred}^{60}, Aux^{30}\right) \tag{6}$$

where $Aux^{30}$ represents the 30 m auxiliary variables; and $T_{pl}^{30}$ denotes the 30 m LST pseudo-label. A downsampled pseudo-label ($T_{de}^{60}$) and 30 m auxiliary variables are input into the second fine-tuned model, formulated as

$$T_{pred}^{30} = Net_{fine2}(T_{de}^{60}, Aux^{30}) \tag{7}$$

where $Net_{fine2}$ is the second fine-tuned model initialized from $Net_{fine1}$, and $T_{pred}^{30}$ is the 30 m downscaled LST. Similarly, the 30 m LST pseudo-label and the fine-resolution auxiliary variable act as supervisory signals. It is worth mentioning that limited by the spatial resolution of Landsat-derived auxiliary variables, this study conducted only two fine-tuning stages.

Theoretically, the proposed framework possesses the potential to further generate LST products at finer spatial resolutions when higher-resolution auxiliary variables are available.

### *3.2 Cross-modal Fusion Downscaling Network for LST (CFDN-LST)*

To effectively exploit the complementary information among coarse-resolution LST and fine-resolution auxiliary variables, we propose the CFDN-LST network. This network retains the proven multi-scale multi-temporal super-resolution reconstruction (MSMTSR) network (Li et al., 2025) while being enhanced by a core learnable dictionary attention mechanism for superior cross-modal feature representation.

As show in Fig. 3, CFDN-LST takes two input streams: coarse-resolution LST and fine-resolution auxiliary variables. The overall architecture comprises five core modules. First, the preliminary feature extraction projects two input modalities into a unified feature space via convolutions; while the auxiliary variables additionally utilize a Multi-scale Feature Extraction (MFE) block (Xiao et al., 2024) to capture multi-scale context. Second, cross-modal fusion is performed by the Cross-modal Dictionary Attention Fusion (CDAF) block (Fig.  (a)) to facilitate intricate feature interactions. Third, deep local features are extracted using a Lightweight Convolutional structure (LCM) featuring the Advanced High-frequency Preserving Block (AHPB) showed in (Fig.  (c)), which is specifically designed to strengthen the modeling of fine-texture features through explicit high-frequency calculation. Fourth, global dependencies are modeled by the Lightweight Transformer Module (LTM) consisting of Efficient Transformer (ET) blocks (Lu et al., 2022). Finally, the texture reconstruction module maps the aggregated features to the target resolution while generates the final fine-resolution LST output.

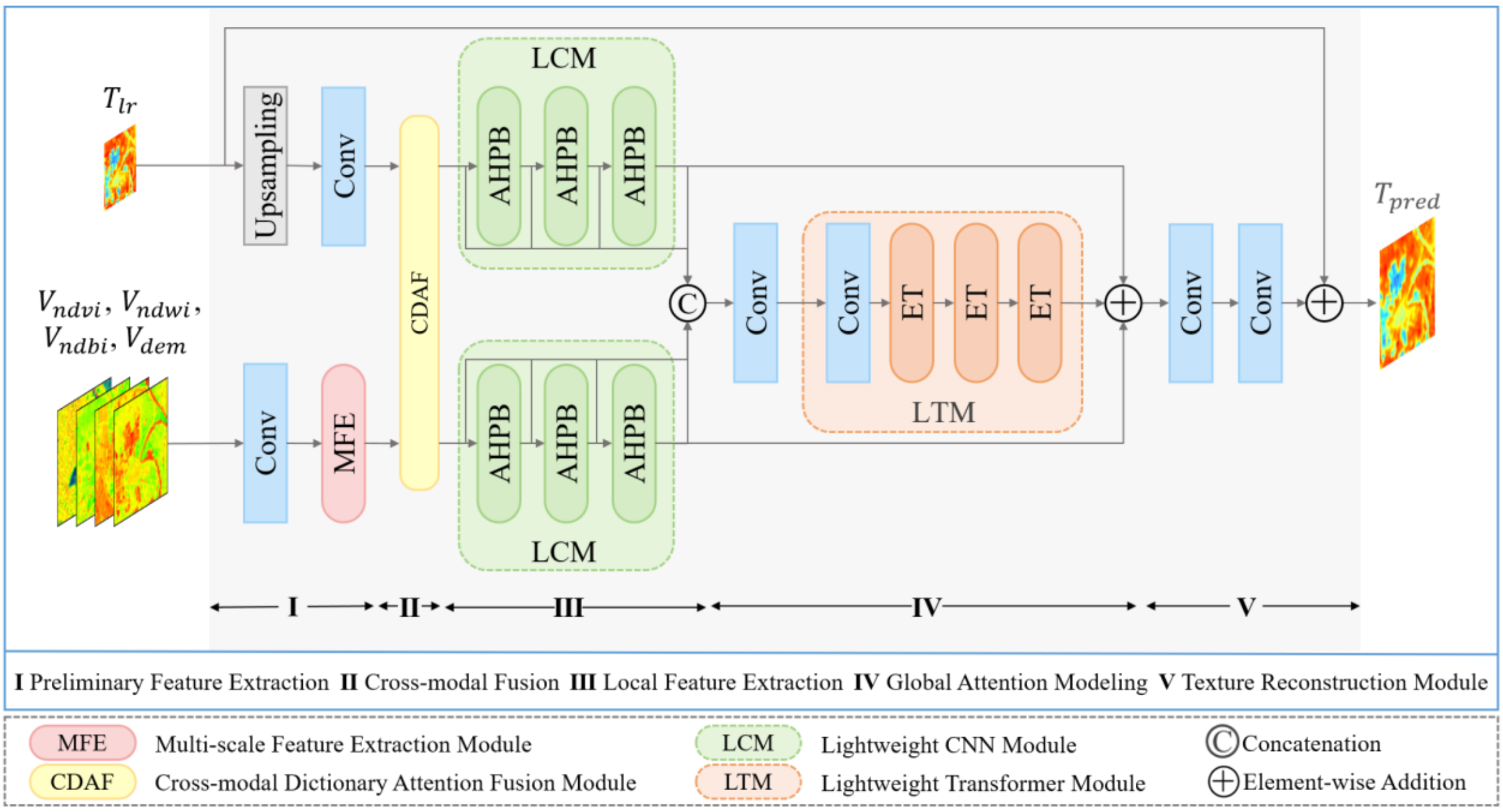

Fig. 3 Main architecture of the CFDN-LST. $T_{lr}$ denotes low resolution LST, V represents auxiliary variables and $T_{pred}$ is the predicted LST.

The core innovation of CFDN-LST lies in a customized CDAF block. Unlike conventional methods that rely on simple concatenation or channel-wise weighting, the CDAF block employs a Learnable Dictionary Attention Representation (LDAR) block (Fig. (a)) to perform deep feature integration. At its heart, LDAR constructs a learnable dictionary to encode and reconstruct input features while leveraging global attention mechanisms to selectively emphasize task-relevant patterns(Li et al., 2025), which enables the network to explicitly distinguish between common characteristics and modality-specific information, producing more structured and interpretable feature representations. Subsequently, by reconstructing and reweighting feature representations, CDAF maximizes the complementary information from fine-resolution auxiliary data to augment the LST features, resulting in a more expressive fused representation.

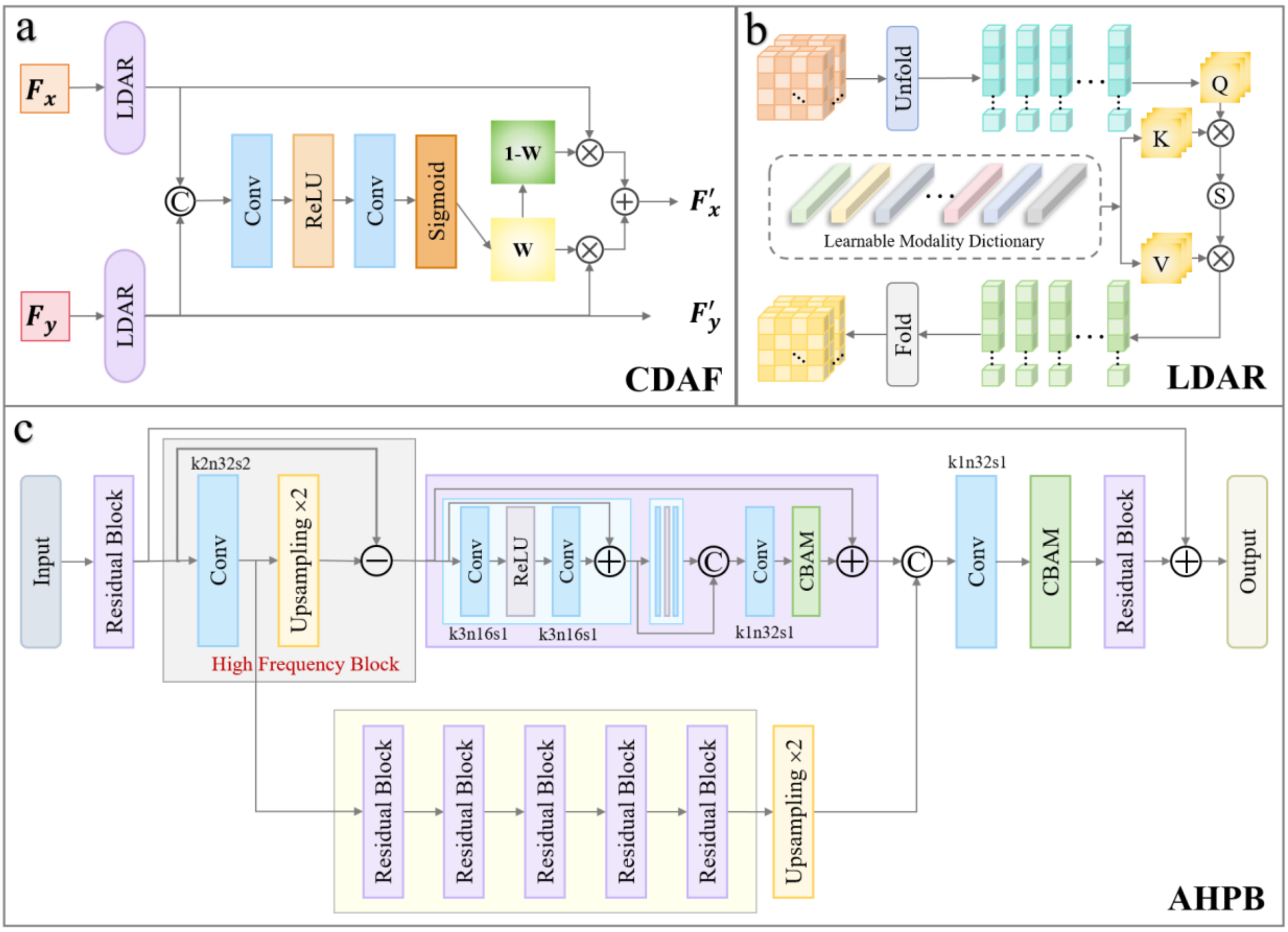

Fig. 4 Schematic of (a) CDAF, (b) LDAR and (c) AHPB. $F_x$ and $F_y$ represent the initial features from two branches and serve as the inputs to CDAF. The outputs are denoted as $F_x'$ and $F_y'$. Query (Q), Key(K), and Value(V) are vectors used in the attention mechanism computation.

### *3.3 Loss Function*

To achieve reliable fine-scale LST reconstruction in the absence of true fine-resolution labels, a composite stage-dependent loss functions is employed throughout the progressive learning framework and it can be expressed as:

$$\mathcal{L}_{total} = \mathcal{L}_{data} + \lambda \cdot \mathcal{L}_{hp} \tag{8}$$

where the total loss $\mathcal{L}_{total}$ of each stage can be described as combination of a data consistency term $\mathcal{L}_{data}$ with a high-frequency structural term $\mathcal{L}_{hp}$.

The data consistency loss $\mathcal{L}_{data}$ computes the L1 distance between the predicted LST and the target label, ensuring pixel-wise accuracy in thermal values:

$$\mathcal{L}_{data} = \frac{1}{W \cdot H} \sum_{x=1}^{W} \sum_{y=1}^{H} \left\| T_{pred}(x, y) - T_{label}(x, y) \right\|_1 \tag{9}$$

where $T_{pred}$ represents the predicted LST; $T_{label}$ denotes the supervisory signal used in each training stage. Specifically, in the pre-training stage, it corresponds to the 120 m Landsat LST; in the two fine-tuning stages, it refers to the LST pseudo-labels generated from the previous stage.

The high-frequency structural loss $\mathcal{L}_{hp}$ leverages structural prior from fine-resolution auxiliary variables to enhance spatial fidelity and compensate for LST pseudo-label uncertainty. Among the available auxiliary variables, NDVI is adopted as the structural reference due to its strong and well-documented correlation with LST (Nemani and Running. 1997, Weng et al., 2004, Mildrexler et al., 2011). Thus, $\mathcal{L}_{hp}$ can be expressed as:

$$\mathcal{L}_{hp} = \frac{1}{W \cdot H} \sum_{x=1}^{W} \sum_{y=1}^{H} \left\| \mathcal{H}(T_{pred}(x, y)) - \mathcal{H}(V_{ndvi}(x, y)) \right\|_1 \tag{10}$$

where $V_{ndvi}$ is the fine-resolution NDVI image in each stage; $\mathcal{H}()$ denotes a Laplacian filter which used to extract high-frequency components.

Moreover, adjustable weight $\lambda$ controls the contribution of the high-frequency structural loss $\mathcal{L}_{hp}$. In the pre-training stage, where real 120 m LST labels are available, the loss is solely governed by data consistency ($\lambda = 0$). In the fine-tuning stages that rely on LST pseudo-labels, $\mathcal{L}_{hp}$is activated to preserve spatial details with $\lambda$ empirically set to 0.1 and 0.2 in the two subsequent stages, respectively.

### *3.4 Experimental Setup*

Following the progressive training strategy described in Section 3.1, experimental datasets were constructed for the three stages. For each stage, paired inputs were generated by cropping coarse-resolution LST and fine-resolution auxiliary variables into 32 × 32 and 64 × 64 patches, respectively, ensuring consistent spatial coverage. The resulting dataset for each stage was randomly split into training, validation, and testing subsets in a ratio of 60% / 20% / 20%. Model training was performed using the Adam optimizer. The initial learning rate was set to $2 \times 10^{-4}$ for pre-training stage and $1 \times 10^{-4}$ for both fine-tuning stages, with a decay factor of 0.5 applied every 50 epochs. The pre-trained model was trained for 250 epochs, whereas each fine-tuning stage was trained for 30 epochs.

## 4. Results and Evaluation

To comprehensively evaluate the effectiveness of the proposed framework, we carry out spatial details and pattern validation via the Chinese Land Cover Dataset (CLCD) (Yang and Huang. 2021) and SDGSAT-1 imagery. Subsequently, in situ observations were used to perform temporal validation of the downscaling LST results. Furthermore, we conducted the performance comparison against existing representative LST downscaling methods. Both reference-based quantitative metrics and reference-free quality indicators were employed to assess the results.

### *4.1 Assessment of Spatial Details and Patterns*

Spatial details and patterns validation of the downscaled results was conducted through a dual approach, using both the land-cover data CLCD to assess internal structure and 30 m SDGSAT-1 LST as an independent benchmark.

Fig. 5 presents the large-area 30 m LST mapping alongside four zoomed-in subregions and corresponding CLCD patterns. Overall, the proposed 30 m downscaled LST captures realistic spatial textures, continuous thermal gradients, and sharper land-cover boundaries across diverse landscapes. In urban areas, it clearly distinguishes impervious surfaces and their details, which indicated by the red box in Fig.  (a); in regions with multiple interwoven features, the model can reconstruct fine roads, rivers, and their internal textures (as shown in the red boxes in Fig.  (b), (c) and (d)). Compared with the heavily smoothed official cubic product, the proposed LST exhibit stronger spatial correspondence with the 30 m CLCD map and more accurately delineate land-cover boundaries and their associated thermal transitions, demonstrating the effectiveness of the method in restoring fine-scale thermal heterogeneity.

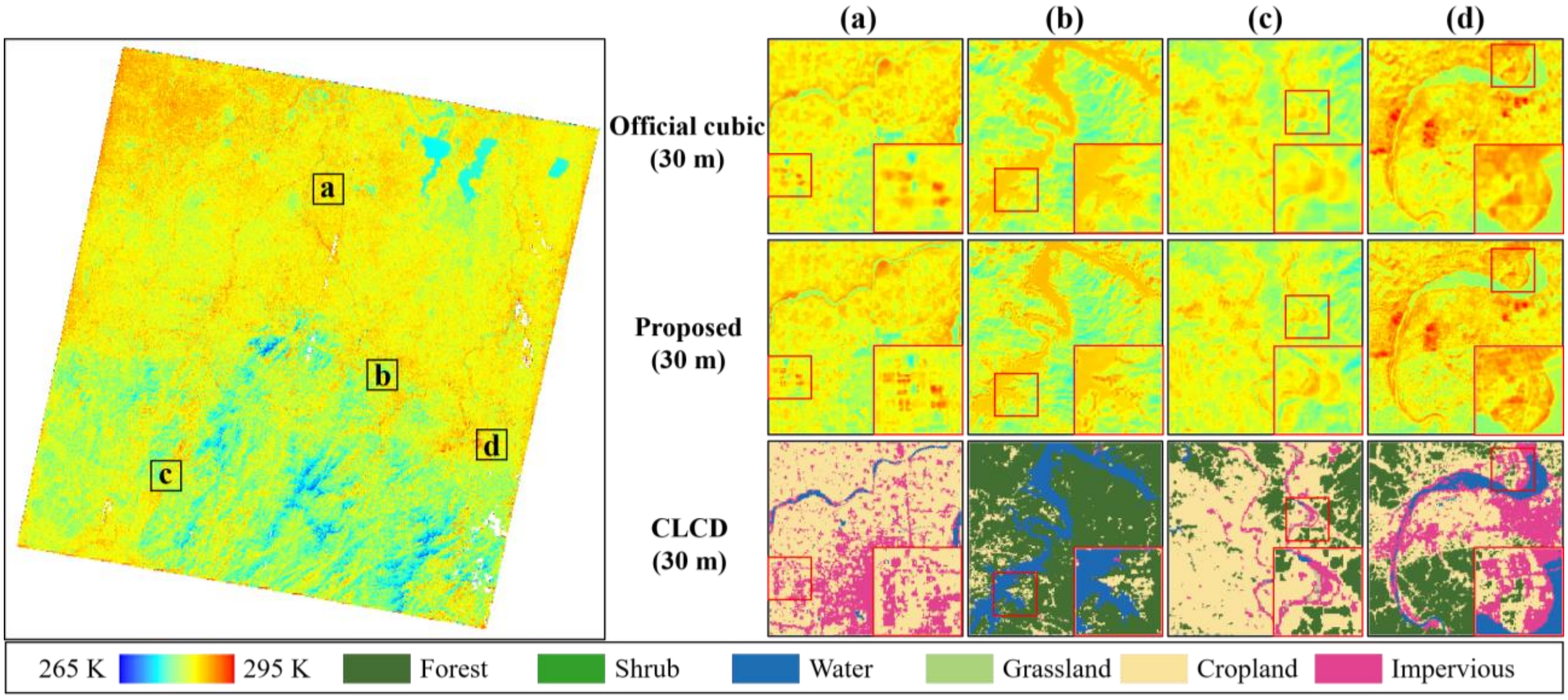

Fig. 5 Large-area mapping of the proposed 30 m LST (left), along with enlarged views of four subregions (right (a)-(d)).

Fig. 6 compares the official cubic LST and the proposed 30 m LST with the SDGSAT-1 30 m LST. Due to inherent differences in sensor specifications, overpass times and retrieval algorithms between Landsat and SDGSAT-1, histogram matching is employed to mitigate systematic biases, while both LST and high-frequency maps are presented for comprehensive comparison. Across three groups, the proposed LST shows faithfully reproduces both general

LST distribution and fine-scale spatial structures. Specifically, in landscape transition zones as Fig. 6 (a), high-frequency maps reveal strong agreement with SDGSAT-1 in fine-scale structural details and land-cover boundaries. For thermally heterogeneous regions in Fig. 6 (b), the proposed LST accurately reproduces warm-cool boundaries and local temperature patches. Fig. 6 (c) shows pronounced topographic variation, both the LST and high-frequency components indicate that the proposed results capture terrain-driven thermal gradients. In contrast, the official cubic product remains overall consistent but exhibits over-smoothing and loss of spatial detail. These findings indicate that the proposed method recovers reliable thermal structures with higher spatial integrity and texture fidelity than official cubic product.

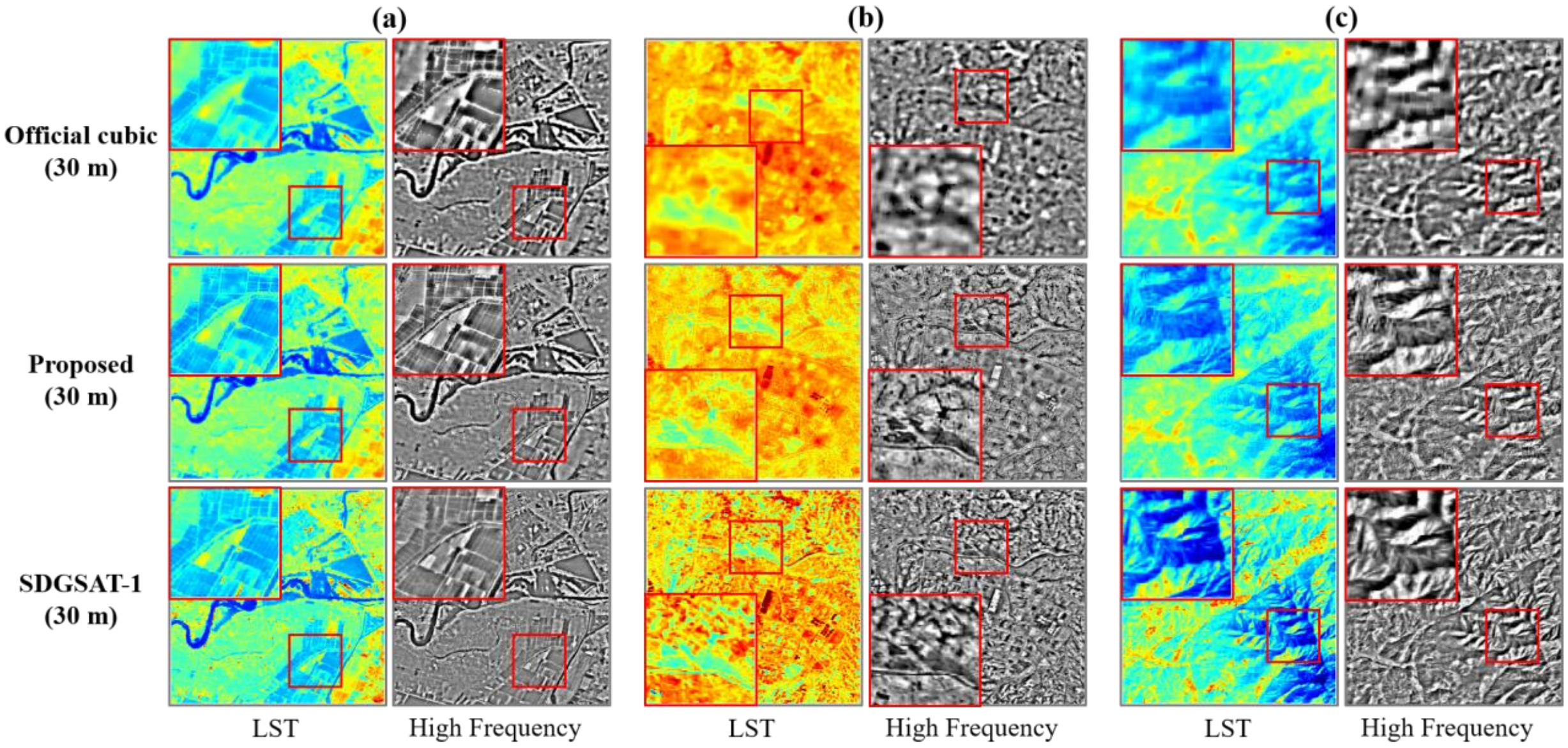

Fig. 6 Comparison of spatial patterns and high-frequency components of the official cubic LST and the proposed downscaled LST against the SDGSAT-1 LST reference. Panels (a–c) correspond to three representative subregions.

These visual improvements are further supported by the density scatter plots and quantitative evaluations in Fig. 7. The scatter centers for both official cubic LST and the proposed LST are closely aligned with the 1:1 diagonal after histogram matching. Notably, the proposed method achieves a near-zero Bias (0.009 K) and the lower RMSE (2.359 K vs. 2.850

K), indicating superior absolute precision. Beyond numerical alignment, we employed SpearR and CrossENT to evaluate informational and structural consistency, and the proposed LST exhibits a markedly lower CrossENT (3.256 vs. 4.041) than the cubic interpolation. These improvements confirm that the proposed framework more effectively preserves complex spatial patterns and thermal heterogeneity, ensuring high-fidelity downscaling results even when validated against independent high-resolution satellite observations.

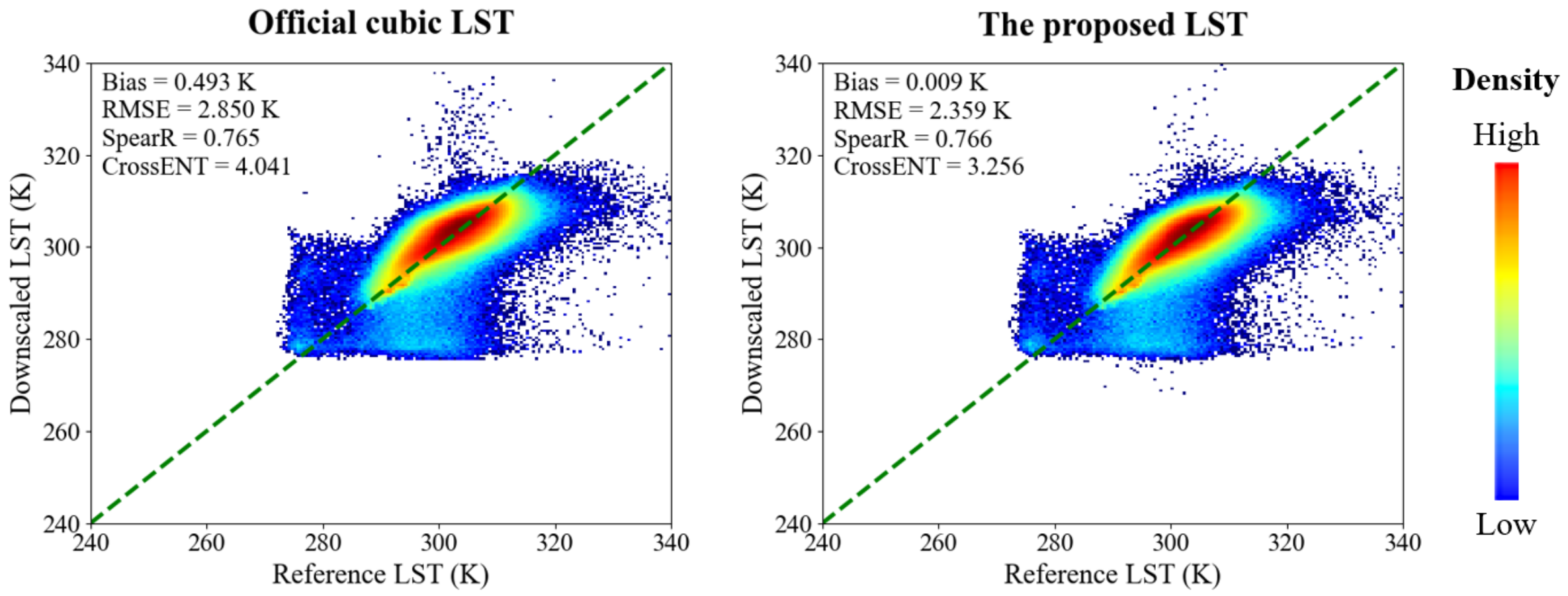

Fig. 7 Density scatter plots of the official cubic and the proposed LST against SDGSAT-1 LST reference. Quantitative assessed using bias, Root Mean Square Error (RMSE), Spearman's Rank Correlation (SpearR) and Cross Entropy (CrossENT).

### *4.2 Validation Against In Situ Measurements*

In situ observations from the Heihe station network were used to further validate the performance of the downscaled LST results. Table  presents the validation results of the official cubic 30 m LST product and the proposed downscaled 30m LST during 2022-2023. Bold values indicate superior performance. As listed in Table , the proposed method achieves superior overall accuracy, with lower mean Bias (0.110 K vs. 0.813 K), MAE (3.353 K vs. 3.710 K), and RMSE (4.052 K vs. 4.411 K), and a higher mean $R^2$ (0.859 vs. 0.831) compared to the official cubic product.

At the site level, the proposed method demonstrates robust performance that outperforming the official cubic product at majority of stations (JYL, HZZ, YK, DM, AR). Slightly lower performance of the proposed method is observed at HH and ZY stations, likely due to the unique surface characteristics of these stations. The HH site (cultivated grassland) is strongly influenced by human activities and soil moisture conditions, while the ZY site (wetlands) exhibits complex hydrothermal processes and pronounced mixed-pixel effects. The official cubic product provides a smoothed "average temperature", whereas the proposed method aims to reveal intra-pixel details and small-scale thermal textures, which may introduce slight discrepancies in such highly heterogeneous environments. Overall, these results validate that the downscaled LST preserves the high-accuracy characteristics of satellite-derived LST while enhancing spatial detail and improving consistency with point-based observations.

**Table 3**

Quantitative evaluation of the 30 m downscaled results using in situ observations

| Site | Data | Bias (K) | MAE (K) | RMSE (K) | $R^2$ |
|---|---|---|---|---|---|
| Idea data | — | 0 | 0 | 0 | 1 |
| AR | Official cubic | 3.856 | 4.429 | 5.132 | 0.640 |
| | Proposed | **2.053** | **3.422** | **4.047** | **0.776** |
| DM | Official cubic | 2.196 | 3.110 | 3.727 | 0.878 |
| | Proposed | **1.491** | **2.513** | **3.139** | **0.913** |
| HH | Official cubic | **-5.279** | **5.279** | **5.941** | **0.849** |
| | Proposed | -5.354 | 5.359 | 6.027 | 0.845 |
| HZZ | Official cubic | 2.108 | 4.273 | 5.199 | 0.905 |
| | Proposed | **0.970** | **3.216** | **4.282** | **0.936** |
| JYL | Official cubic | 2.139 | **5.158** | **5.943** | **0.637** |
| | Proposed | **1.865** | 5.169 | 5.966 | 0.634 |
| YK | Official cubic | 1.735 | 2.010 | 2.450 | 0.958 |
| | Proposed | **1.078** | **1.904** | **2.210** | **0.965** |
| ZY | Official | **-1.065** | **1.710** | **2.483** | **0.951** |
| | cubic | | | | |
| | Proposed | -1.333 | 1.887 | 2.690 | 0.943 |
| Average | Official cubic | 0.813 | 3.710 | 4.411 | 0.831 |
| | Proposed | **0.110** | **3.353** | **4.052** | **0.859** |

*Note. RMSE and R2 represent root mean square error and coefficient of determination respectively.

*4.3 Performance Comparison over Existing Downscaling Methods*

To evaluate the efficacy of the proposed method, we compared it with two typical LST downscaling approaches: the linear-regression-based Thermal sHARPening (TsHARP) (Agam et al., 2007) and the deep-learning-based Land Surface Temperature Downscaling Residual Network (LSTDRN) (Zhang et al., 2021). Both were strictly reproduced by following their original designs ensure a fair comparison. As shown in Fig. 8, all three downscaling methods further improve spatial detail beyond the official cubic product, demonstrating the validity of downscaling. Among them, TsHARP produces spatial discontinuities (Fig. 8 (b), (c)) and noisy textures (Fig. 8 (a)). This instability stems from its reliance on simplified linear regression, which fails to capture complex heterogeneity and is prone to over-extrapolation. LSTDRN preserves large-scale LST patterns but textures tend to be over-smoothed, with many fine-scale details lost. Such limitations arise because its coarse-resolution feature learning without scale alignment or high-frequency guidance. In contrast, the proposed method substantially enhances spatial details while maintaining radiometric stability. This is achieved through a tailored cross-modal fusion network that mines nonlinear complementary information, alongside a progressive self-training strategy to ensure scale-consistent mappings.

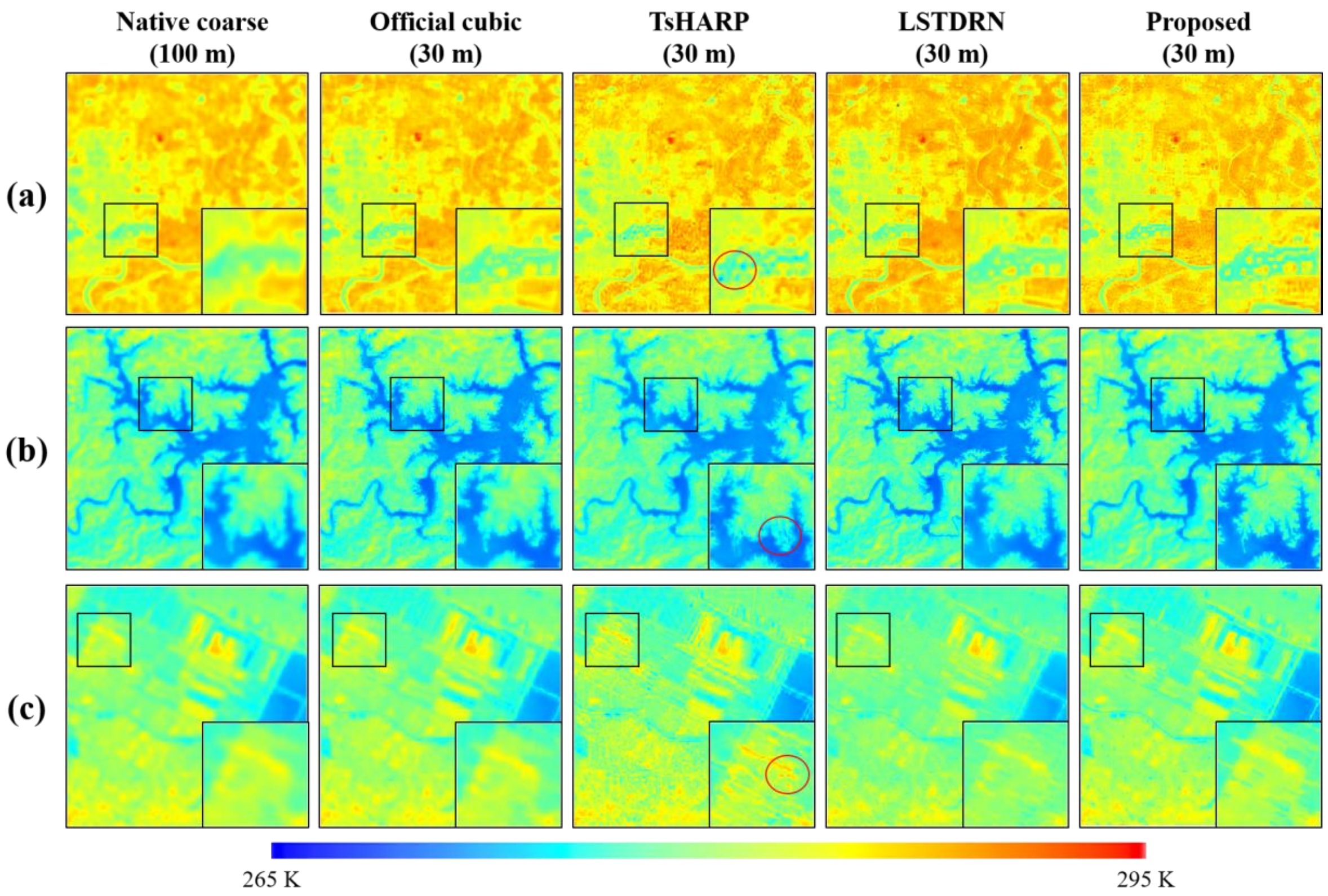

Fig. 8 Downscaling results of different methods. (a)–(c) correspond to results from three representative subregions.

Quantitatively, the simulation experiment was conducted on downsampled data across diverse land cover types, using the available Landsat native coarse LST as a reference to verify algorithm performance, with results showed in the radar charts (Fig. 9). From an overall perspective, the proposed method achieves the highest global accuracy compared to TsHARP and LSTDRN, which represents the aggregate statistics across all land cover categories combined. Moreover, the proposed approach maintains the smallest error (MAE and RMSE) and the largest R2 coverage the vast majority of individual land cover types, demonstrating its superior capability to handle surface heterogeneity. Additionally, for the 30 m downscaling results of real experiment, a hybrid evaluation was performed by incorporating both reference-based and no-reference metrics due to the absence of ground-truth; the results are listed in Table 4, where bold and underline indicate the best and second-best performance, respectively.

Specifically, the reference-based evaluation involved downsampling the downscaled results to coarse resolution to assess thermal consistency with coarse observations, while the reference-free assessment was conducted directly at 30 m to quantify spatial detail enhancement. As shown in the figure and table, the proposed method achieves the highest accuracy across all indicators in the two experimental scenarios. The simulation results demonstrate the superiority of the proposed cross-modal fusion network, while the evaluation of 30 m downscaling results further validate the effectiveness of the proposed progressive self-training framework.

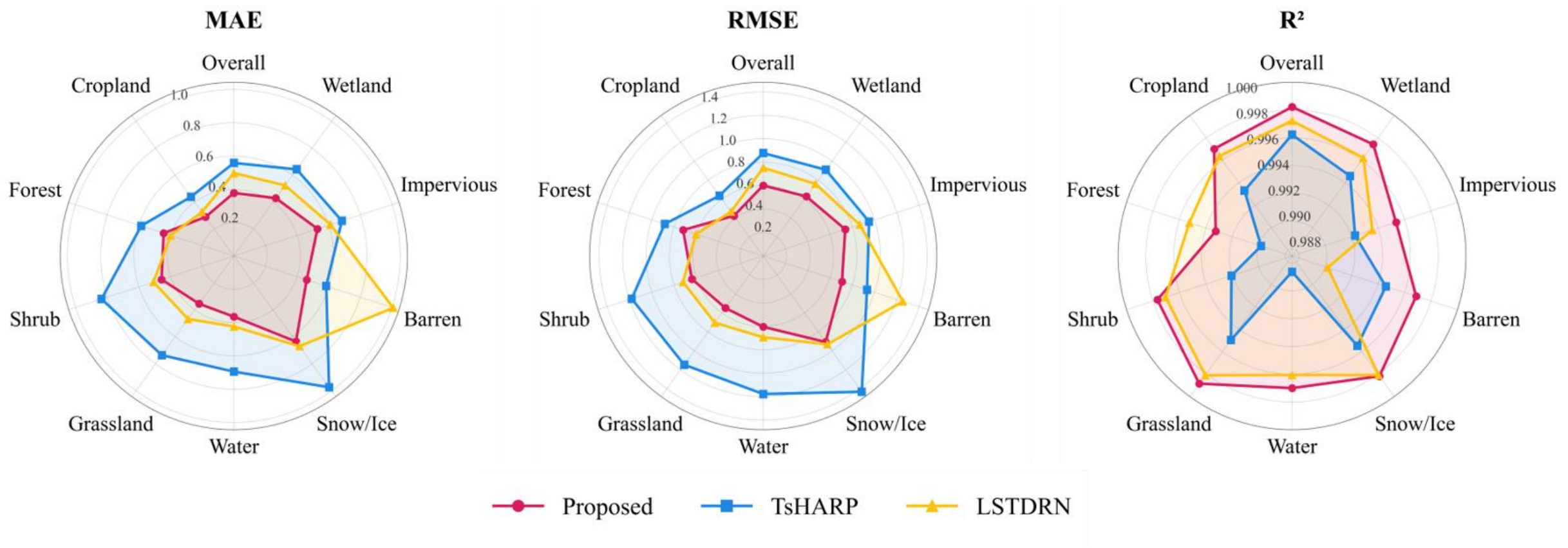

Fig. 9 Radar charts of downscaling performance comparison for different land cover categories in the simulation experiment. "Overall" represents the aggregate statistics across all land cover categories combined.

**Table 4**

Quantitative evaluation of the real experiment

| Method\ Merics | reference-based evaluation | | | reference-free evaluation | | |
|---|---|---|---|---|---|---|
| | MAE (K) | RMSE (K) | $R^2$ | AG (K) | SF (K) | NIQE |
| Ideal Data | 0 | 0 | 1 | +∞ | +∞ | 0 |
| Official cubic | — | — | — | 0.349 | 6.610 | 9.076 |
| TSHARP | 0.546 | 0.754 | 0.899 | 0.400 | 8.761 | 8.384 |
| LSTDRN | <u>0.536</u> | <u>0.707</u> | <u>0.918</u> | <u>0.401</u> | <u>8.841</u> | <u>8.137</u> |
| Proposed | **0.451** | **0.665** | **0.924** | **0.490** | **9.167** | **6.522** |

*Note. AG, SF, and NIQE represent average gradient, spatial frequency, and natural image quality evaluator, respectively.

## 5. Discussions

We conducted a systematic analysis of the proposed method from three perspectives: reliability of the progressive self-training downscaling strategy through multi-stage experiments, contributions of the input variables and network architecture, and strengths and limitations analysis.

### *5.1 Multi-stage Validation of Progressive Self-training Downscaling Strategy*

As shown in Fig. 10, the progressive evolution of LST patterns across three stages reveals the framework's capability in detailed feature recovery. The initial 120 m results from the pre-training stage capture a thermal distribution consistent with the official cubic products, albeit with limited spatial detail. During the subsequent fine-tuning stages, although the LST pseudo-labels capture the general thermal layout, they inherently suffer from blurred textures (Fig 10 (a), (c)) and unclear boundaries (Fig 10 (b), (d)) due to training-application scale inconsistency. However, the fine-tuned results at the respective resolutions effectively enhance spatial details without compromising radiometric fidelity, which is quantitatively substantiated by the consistent improvements in both reference-based and reference-free metrics in Table 5. This improvement primarily stems from the fine-tuning process enabling the network to learn fine-resolution feature distributions consistent with the application scale and mitigating errors inherited from LST pseudo-labels through high-frequency structural constraints.

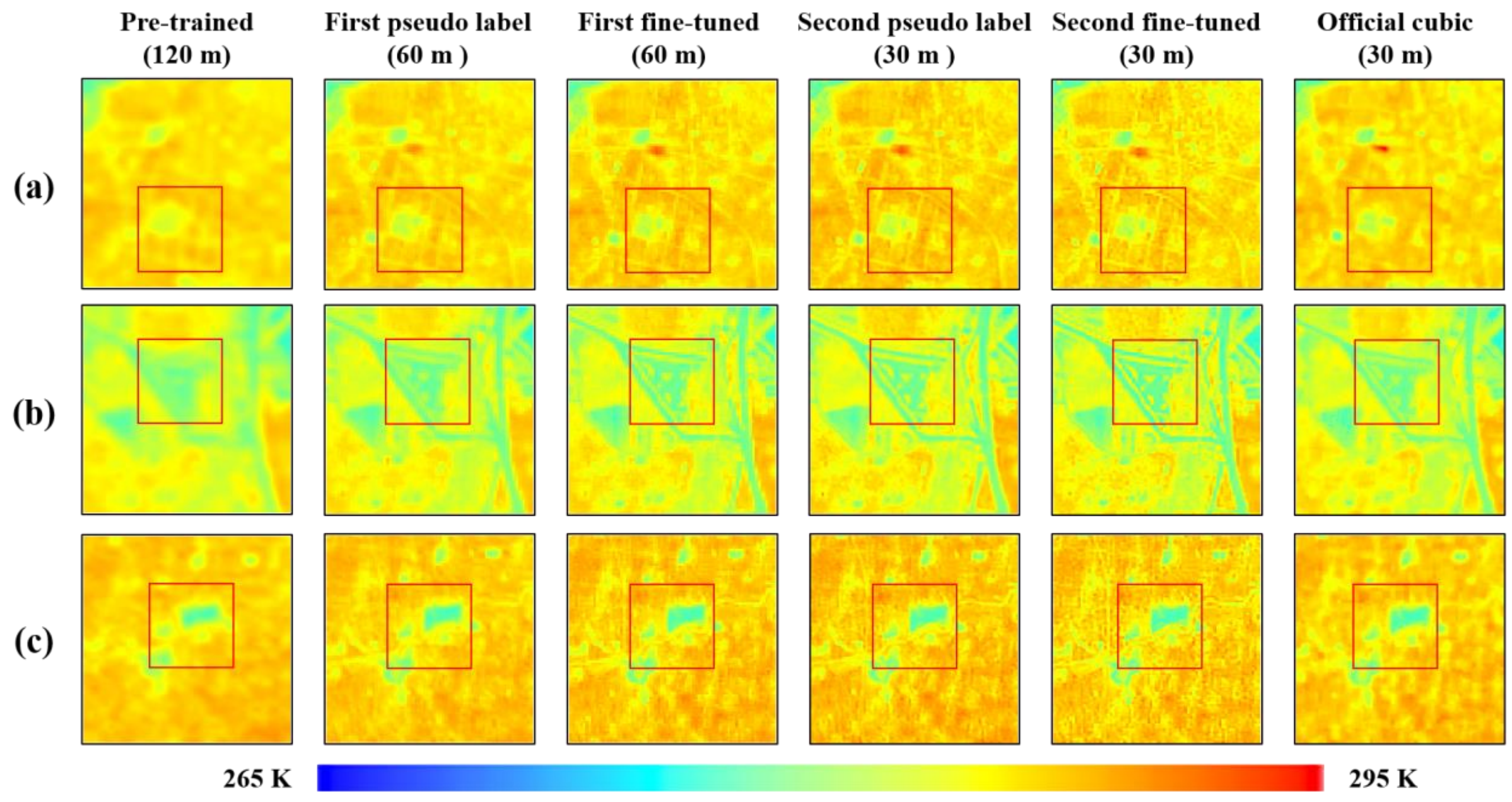

Fig. 10 Visual evolution of downscaled LST maps across the three progressive stages. (a)–(c) represent three different subregions.

**Table 5**

Quantitative assessment of the fine-tuning stages

| **Result \ Metrics** | reference-based evaluation | | | reference-free evaluation | | |
|---|---|---|---|---|---|---|
| | MAE (K) | RMSE (K) | $R^2$ | AG (K) | SF (K) | NIQE |
| Ideal data | 0 | 0 | 1 | +∞ | +∞ | 0 |
| First pseudo label | 0.262 | 0.347 | 0.926 | 0.456 | 0.894 | 13.063 |
| First fine-tuned | **0.212** | **0.285** | **0.935** | **0.527** | **1.103** | **12.285** |
| Second pseudo label | 0.633 | 0.716 | 0.829 | 0.458 | 9.151 | 7.137 |
| Second fine-tuned | **0.445** | **0.569** | **0.831** | **0.490** | **9.167** | **6.522** |

*5.2 Contributions of Auxiliary Variables and Network Components*

(1) Ablation of auxiliary surface variables

Fine-resolution auxiliary variables are essential to the proposed framework. Their contributions were evaluated by individually omitting NDVI, NDWI, NDBI, and DEM on the 240-120 m downscaling dataset. Quantitative results in Table 6 demonstrate that omitting any variable degrades performance, with larger impacts from NDBI, NDVI, and DEM and a smaller effect from NDWI, while both NDWI and DEM notably increase bias. Specifically, removing DEM leads to the loss of terrain-driven thermal gradients, which are indispensable for characterizing the significant vertical temperature variations in mountainous areas. Omitting NDVI and NDWI hinders the model's capacity to jointly constrain surface thermal heterogeneity, as these predictors are fundamental for capturing the vegetation and moisture status. Furthermore, removing NDBI eliminates the critical information needed to distinguish anthropogenic materials from natural backgrounds. Overall, the synergistic use of auxiliary variables enables the model to preserve thermal patterns and spatial structure across different thermal-process dimensions, resulting in more stable and reasonable LST downscaling.

**Table 6**

Quantitative evaluation of the surface variables ablation experiments

| Surface Variables | | | | Metrics | | | |
|---|---|---|---|---|---|---|---|
| DEM | NDVI | NDWI | NDBI | Bias (K) | MAE (K) | RMSE (K) | $R^2$ |
| | √ | √ | √ | 0.026 | 0.549 | 0.721 | 0.736 |
| √ | | √ | √ | **0.001** | 0.547 | 0.725 | 0.738 |
| √ | √ | | √ | 0.016 | <u>0.526</u> | <u>0.691</u> | <u>0.759</u> |
| √ | √ | √ | | **0.001** | 0.565 | 0.742 | 0.729 |
| √ | √ | √ | √ | <u>0.002</u> | **0.520** | **0.683** | **0.765** |

(2) Ablation of core network components

We further assessed the roles of key components (CDAF, MEF, LCM, and LTM) via ablation experiments on the 240-120 m downscaling dataset. Quantitative results are shown in Table 7. Overall, removing any module degrades performance, confirming the necessity of the proposed network design. Omitting the CDAF module causes the largest accuracy drop due to the loss of cross-modal interaction. As the fusion unit linking thermal and spectral features, CDAF bridges their semantic gap and enables complementary feature exploitation; thus without it, cross-modal associations are weakened. Within the network backbone, LCM and LTM exhibit strong complementarity, LCM extracts local features via convolution, and its removal weakens fine-scale LST characterization (higher MAE and RMSE); meanwhile, LTM captures global dependencies through self-attention, and its absence disrupts global thermal coherence (lower PSNR and SSIM). Together, they enable the hybrid CNN-Transformer architecture to jointly reconstruct local details and global structures.

**Table 7**

Quantitative evaluation of components ablation experiments

| Components | | | | Metrics | | | |
|---|---|---|---|---|---|---|---|
| CDAF | MEF | LCM | LTM | Bias (K) | MAE (K) | RMSE (K) | $R^2$ |
| | √ | √ | √ | -0.031 | 0.535 | 0.702 | 0.754 |
| √ | | √ | √ | 0.005 | 0.532 | 0.688 | 0.762 |
| √ | √ | | √ | -0.006 | 0.526 | 0.691 | 0.759 |
| √ | √ | √ | | 0.017 | 0.529 | 0.696 | 0.756 |
| √ | √ | √ | √ | **0.002** | **0.520** | **0.683** | **0.765** |

*5.3 Strengths and Limitations*

The proposed approach leverages a cross-modal fusion network to exploit complementary features of LST and auxiliary variables, and adopts a progressive self-training strategy to learn scale-consistent representations across different resolutions. In addition, the NDVI high-frequency structural guidance effectively mitigates error accumulation caused by LST pseudo-labels. Without relying on fine-resolution ground truth, the framework achieves progressive downscaling of Landsat 8/9 LST from 120m to 30 m, and demonstrates strong stability and effectiveness.

Nevertheless, certain limitations remain. On the one hand, scale effects cannot be fully eliminated in the progressive self-training strategy. The reliance on LST pseudo-labels inevitably introduces uncertainty, and local errors may propagate and accumulate across successive fine-tuning stages, particularly in highly heterogeneous landscapes. As the number of fine-tuning stages increases, such accumulated uncertainties may further amplify. On the other hand, the progressive multi-stage training paradigm entails increased implementation complexity, as model training and inference must be conducted through a sequence of interdependent resolution levels rather than a single unified process.

## 6. Conclusion

In this paper, we developed a progressive self-training framework tailored for downscaling Landsat LST through cross-modal fusion, generating 30 m Landsat LST without relying on fine-scale ground truth while maintaining minimal data dependence. By adopting a pseudo-label-based multi-stage learning strategy, the framework progressively refines fine-scale thermal patterns. The cross-modal fusion network combines dictionary-attention mechanism with the CNN-Transformer architecture to exploit complementary cross-modal information, while embedded a dual frequency-domain guidance to preserve spatial details and suppress artifacts. Validation using SDGSAT-1 LST and in situ observations demonstrates the spatial and temporal reliability of the proposed method. Comprehensive experiments demonstrate that the framework outperforms existing approaches in both spatial detail reconstruction and thermal-pattern preservation, and performs robustly across multiple resolutions. Overall, this framework provides a stable pathway to enhance Landsat LST resolution for fine-scale surface process studies and localized environmental monitoring.

In future, we will focus on optimizing the proposed framework to address the potential error accumulation and the procedural complexity introduced by the multi-stage strategy, with the goal of developing a unified end-to-end paradigm for the generation of fine-resolution and long-term Landsat products. In addition, by further incorporating high-temporal-resolution LST products, we expect to enhance the temporal density of Landsat LST observation, which facilitates refined monitoring of rapid surface thermal dynamics.

## CRediT authorship contribution statement

**Huanfeng Shen:** Conceptualization, Validation, Supervision, Funding acquisition, Writing – review & editing. **Chan Li:** Methodology, Investigation, Formal analysis, Data curation, Writing – original draft, Writing – review & editing. **Menghui Jiang:** Conceptualization, Supervision, Methodology, Funding acquisition, Writing – review & editing. **Penghai Wu:** Supervision, Project administration. **Guanhao Zhang:** Formal analysis, Visualization. **Tian Xie:** Conceptualization, Validation.

## Declaration of competing interest

The authors declare no competing interests.

## Data availability

All data used in this study are publicly available: Landsat 8/9 data and DEM dat a (https://earthexplorer.usgs.gov/). Heihe River Basin site data: (https://data.tpdc.ac.cn/).

## Acknowledgement

The numerical calculations in this paper have been done on the supercomputing system in the Supercomputing Center of Wuhan University. This research was supported in part by the Key Project of the National Natural Science Foundation of China [Grant number 42130108], the National Natural Science Foundation of China [Grant numbers 42571435 and 42301531].